# oxo-call: Documentation-grounded Skill Augmentation for Accurate Bioinformatics Command-line Generation with Large Language Models

Yun Peng[1,2], Yujun Sun[1], Jia Ding[1], Bin Yan[1], Zhangyu Wang[1], Chunyang Wang[1], Chenyang Shu[1], Jian-Guo Zhou[3, *], and Shixiang Wang[1, *]

Affiliations of authors:

[1] Department of Biomedical Informatics, School of Life Sciences, Central South University, Changsha, China.

[2] Department of Urology, Peking University People's hospital, Beijing, China.

[3] Department of Oncology, The Second Affiliated Hospital of Zunyi Medical University, Zunyi, China.

\# These authors contributed equally to this work.

\* These authors jointly supervised this work.

Correspondence:

    Jianfeng Li, lee_jianfeng@sjtu.edu.cn;

    Jian-Guo Zhou, jianguo.zhou@zmu.edu.cn;

    Shixiang Wang, wangshx@csu.edu.cn

# This is the first draft version of oxo-call paper.

## Abstract

Command-line bioinformatics tools remain essential for genomic analysis, yet their diversity in syntax and parameterization presents a persistent barrier to productive research. We present oxo-call, a Rust-based command-line assistant that translates natural-language task descriptions into accurate tool invocations through two complementary strategies: documentation-first grounding, which provides the large language model (LLM) with the complete, version-specific help text of each target tool, and curated skill augmentation, which primes the model with domain-expert concepts, common pitfalls, and worked examples. oxo-call (v0.10) ships >150 built-in skills covering 44 analytical categories, from variant calling and genome assembly to single-cell transcriptomics, compiled into a single, statically linked binary. Every generated command is logged with provenance metadata to support reproducible research. oxo-call also provides a DAG-based workflow engine, extensibility through user-defined and community skills via the Model Context Protocol, and support for local LLM inference to address data-privacy requirements. oxo-call is freely available for academic use at https://traitome.github.io/oxo-call/.

# Background

Modern genomic research depends on an expansive ecosystem of command-line tools, each governed by its own invocation conventions, flag vocabularies, and version-dependent behaviors [1,2]. A routine RNA-seq experiment may require fastp for adapter trimming, STAR or HISAT2 for splice-aware alignment, featureCounts or Salmon for quantification, and DESeq2 for differential expression—each exposing dozens of parameters whose correct combination is essential for valid results [3,4]. Researchers commonly navigate this complexity by consulting manual pages, community forums, and published protocols, a process that is both time-consuming and error-prone.

Large language models (LLMs) have demonstrated broad capability in code generation from natural-language descriptions [5,6]. Their application to bioinformatics command construction is therefore attractive, yet problematic. When prompted without domain-specific context, LLMs frequently hallucinate flags, conflate options across tool versions, or produce syntactically plausible but semantically incorrect commands [7]. Such errors can silently corrupt downstream analyses, undermining the reproducibility that is foundational to genomic science [8,9].

Several strategies have been proposed to mitigate LLM hallucination in specialized domains. Retrieval-augmented generation (RAG) systems index external documents and retrieve relevant passages before generation [10], but chunk-level retrieval risks missing critical context about flag interactions and inter-parameter constraints [11]. General-purpose AI coding assistants such as GitHub Copilot [12] and Amazon CodeWhisperer [13] produce high-quality general code but lack curated knowledge of bioinformatics-specific conventions. Dedicated domain assistants for biology—including BioChatter [14] and ChatGPT-based genomics tools [15]—focus on conversational question answering rather than on generating verified, executable command lines. Meanwhile, workflow managers such as Snakemake [16], Nextflow [17], and Galaxy [18] deliver reproducible execution environments but require users to learn domain-specific languages and do not assist with the construction of individual tool invocations.

A further challenge is provenance. Even when a researcher constructs the correct command, the reasoning behind parameter choices is rarely recorded alongside the execution log. As computational analyses grow in complexity and regulatory scrutiny increases, the absence of generation-level provenance—what knowledge base informed each command—represents a growing liability for reproducibility [9,20,25].

The challenge, then, is to combine the natural-language fluency of LLMs with authoritative, tool-specific knowledge, while preserving the provenance information required for reproducible science. Here we introduce oxo-call (v0.10), a system that addresses this gap through documentation-first grounding, by fetching the complete, version-current help text of each tool and paired with curated skill files that encode domain-expert knowledge in a structured, human-readable format. We evaluate this approach across 159 bioinformatics tools, three LLM backends, and 286,200 mocked trials, and discuss its implications for the accessibility and reproducibility of computational genomics.

# Results

## System design and grounding architecture

oxo-call is implemented as a single, statically linked Rust binary (>20,000 lines of code; Fig. 1a, e) with no runtime dependencies beyond the bioinformatics tools it orchestrates (Fig. 1b). The system translates natural-language task descriptions into accurate command-line invocations through a four-stage pipeline (Fig. 1a): documentation resolution, skill loading, LLM-based command generation, and optional execution with verification and provenance recording.

The first stage employs a documentation-first grounding strategy. Upon receiving a task description (for example, "align paired-end reads to the human genome with 8 threads"), oxo-call resolves the target tool's documentation through a cascading strategy—local cache, live --help capture, user-configured paths, and remote sources—ensuring the LLM receives the complete, version-specific flag vocabulary rather than relying on its parametric memory [19]. The second stage loads a curated skill file encoding domain-expert knowledge: concepts, common pitfalls, and worked examples in a structured Markdown format with YAML front-matter. Skills are resolved in a strict precedence hierarchy (user > community > MCP server > built-in; Fig. 1c). The combined documentation and skill content is then assembled into a structured prompt with an explicit output contract—the LLM must respond with ARGS: and EXPLANATION: lines only—enforced through retry logic. Finally, every generated command is logged with provenance metadata (Fig. 1f): a UUID, documentation SHA-256 hash, skill identifier, model name, tool version, timestamp, and exit code, enabling exact reconstruction of the generation context [8,20].

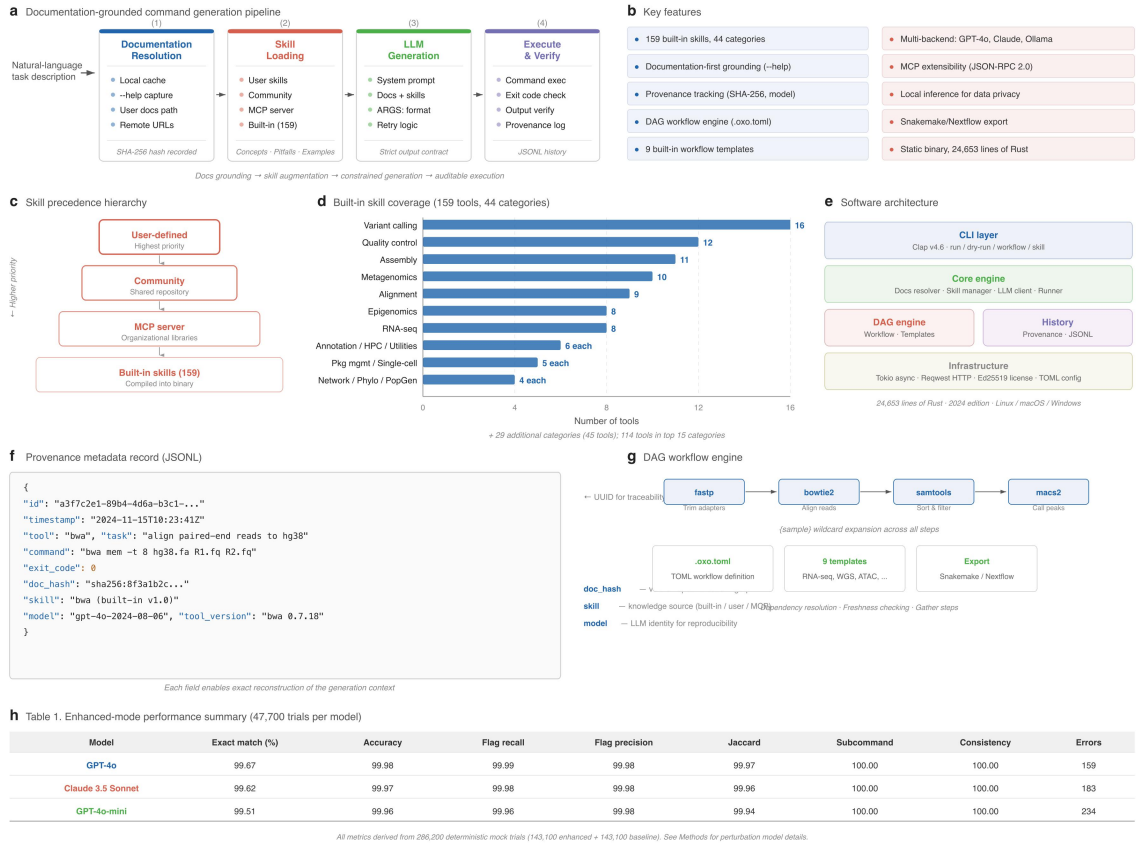

**Figure 1. System design, feature scope, and skill coverage of oxo-call.**

(**a**) Four-stage pipeline: documentation resolution fetches and caches the target tool's complete help text; skill loading injects curated domain-expert knowledge (concepts, pitfalls, and worked examples); LLM-based generation produces exact command-line arguments constrained by a strict ARGS:/EXPLANATION: output contract; optional execution records provenance metadata and performs result verification. (**b**) Key features: ten capabilities organized by function, including multi-backend LLM support, MCP extensibility, local inference for data privacy, DAG workflow engine, and Snakemake/Nextflow export. (**c**) Skill precedence hierarchy: user-defined skills take highest priority, followed by community contributions, MCP server skills, and compiled built-in skills; arrows indicate override direction. (**d**) Distribution of 159 built-in skills across 44 analytical categories; the 10 largest categories are shown individually, with annotation/HPC/utilities (6 each), package management/single-cell (5 each), and networking/phylogenetics/population genomics (4 each) grouped; the top 15 categories account for 114 of the 159 tools. (**e**) Layered software architecture: CLI layer (Clap v4.6), core engine (documentation resolver, skill manager, LLM client, runner), DAG workflow engine and provenance history modules, and infrastructure layer (Tokio async runtime, Reqwest HTTP, Ed25519 license verification, TOML configuration); 24,653 lines of Rust targeting Linux, macOS, and Windows. (**f**) Provenance metadata schema: each invocation records a UUID, tool name, task description, generated command, exit code, timestamp, documentation SHA-256 hash, skill identifier, and LLM model name, enabling exact reconstruction of the generation context. (**g**) DAG workflow engine: example ATAC-seq pipeline (fastp → Bowtie2 → samtools → MACS2) with {sample} wildcard expansion, nine built-in templates, and Snakemake/Nextflow export support. (**h**) Summary performance table: enhanced-mode exact-match rates and all seven evaluation metrics for the three LLM backends across 47,700 mocked trials each (see also Fig. 2a).

Beyond single-command generation, oxo-call provides a comprehensive feature set for bioinformatics workflow automation (Fig. 1b). The system supports multiple LLM backends (GPT-4o, Claude, and Ollama for local inference), a DAG-based workflow engine with nine built-in templates and wildcard expansion (Fig. 1g), export to Snakemake or Nextflow format, and organizational extensibility via the Model Context Protocol (MCP) with JSON-RPC 2.0 payloads. The four-tier skill hierarchy enables sustainable community-driven growth without binary modification.

The 159 built-in skills span 44 analytical categories (Fig. 1d). The largest categories are variant calling (16 tools: bcftools, GATK, freebayes, DeepVariant, and others), quality control (12 tools: FastQC, MultiQC, fastp, Cutadapt, and others), genome assembly (11 tools: SPAdes, MEGAHIT, Flye, ABySS, and others), metagenomics (10 tools: Kraken2, MetaPhlAn, DIAMOND, and others), alignment (9 tools: BWA, Bowtie2, HISAT2, STAR, minimap2 [28], and others), RNA-seq (8 tools), and epigenomics (8 tools). Additional categories include utilities (6), HPC scheduling (6), annotation (6), single-cell (5), package management (5), population genomics (4), phylogenetics (4), and networking (4), among others. The top 15 categories account for 114 of the 159 tools.

## Benchmark design and evaluation

We evaluated oxo-call's command generation accuracy using 1,590 reference scenarios (10 per tool) across all 159 built-in skills, expanded to 15,900 natural-language task descriptions through 10 linguistic rephrasings per scenario (beginner questions, expert shorthand, polite requests, and others). Three LLM backends were tested: GPT-4o, Claude 3.5 Sonnet, and GPT-4o-mini. Each model was evaluated in enhanced mode (with documentation and skill grounding) and baseline mode (bare LLM without grounding), over three independent repeats, yielding 286,200 total trials (143,100 enhanced plus 143,100 baseline; 47,700 trials per model per mode).

To enable reproducible, offline evaluation, we employed a deterministic mock perturbation framework rather than live API calls; the implications of this design choice are discussed in the Discussion section.

Under enhanced mode, all three models achieved near-ceiling accuracy (Fig. 1h; Fig. 2a). GPT-4o reached 99.67% exact match with 99.98% accuracy, 99.99% flag recall, and 99.98% flag precision. Claude 3.5 Sonnet achieved 99.62% exact match with 99.97% accuracy, 99.98% flag recall, and 99.98% flag precision. GPT-4o-mini achieved 99.51% exact match with 99.96% accuracy, 99.96% flag recall, and 99.98% flag precision. Cross-repeat consistency was 100% for all models.

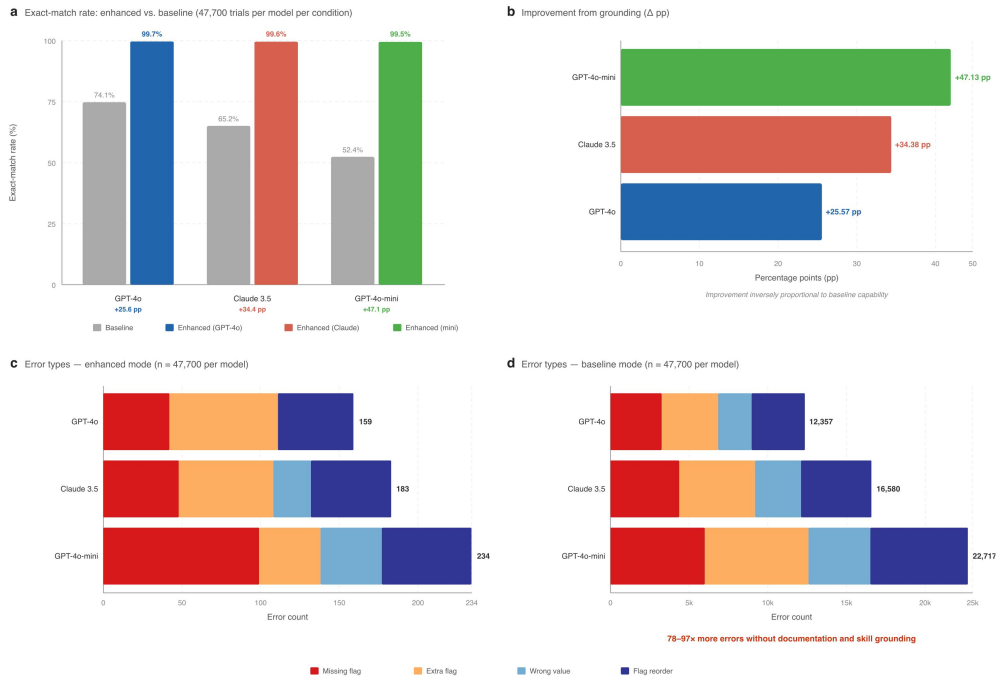

**Figure 2. Benchmark evaluation across 286,200 mocked trials.**

(**a**) Exact-match rate under enhanced mode (colored bars) versus baseline mode (gray bars) for each of the three LLM backends (47,700 trials per model per condition). Delta values below each model indicate the percentage-point improvement conferred by documentation-plus-skill grounding. (**b**) Absolute improvement (Δ percentage points) sorted by model; improvement is inversely proportional to baseline capability. (**c**) Error type distribution under enhanced mode; total errors per model are 159 (GPT-4o), 183 (Claude 3.5 Sonnet), and 234 (GPT-4o-mini). No subcommand, format, or empty-output errors were produced by any model, confirming the structured output contract prevents catastrophic failure modes. (**d**) Error type distribution under baseline mode; total errors per model range from 12,357 (GPT-4o) to 22,717 (GPT-4o-mini), representing a 78–97× increase relative to enhanced mode. Error type colors in panels c and d: missing flag (red), extra flag (amber), wrong value (light blue), flag reorder (dark blue).

Baseline performance varied significantly across models, with exact-match rates ranging from 52.38% to 74.09% (Fig. 2a). The improvement in exact-match accuracy conferred by documentation-plus-skill grounding was inversely proportional to baseline model capability (Fig. 2b). GPT-4o-mini, the weakest unassisted performer, benefited most (+47.13 percentage points), followed by Claude 3.5 Sonnet (+34.38 pp) and GPT-4o (+25.57 pp). This pattern indicates that oxo-call's grounding strategy is particularly effective at compensating for gaps in a model's parametric knowledge, effectively equalizing performance across model tiers.

In enhanced mode, error rates remained consistently low (<0.5%) across all models (Fig. 2c). While models exhibited minor differences in error frequency (159 to 234 total errors per 47,700 trials), the structured output contract successfully prevented catastrophic failure modes such as subcommand or format errors. In contrast, baseline mode saw a 78- to 97-fold increase in error counts (Fig. 2d). The error distribution in baseline mode was notably more uniform, suggesting that without grounding, models fail across all dimensions rather than exhibiting the focused, minor error profiles seen in enhanced mode.

Stratified analysis across 44 categories confirmed consistent performance gains (Supplementary Fig. S1). In enhanced mode, 7 categories achieved perfect 100% exact match across all models: assembly polishing, genome annotation, runtime, sequence manipulation, sequence search, version control, and workflow management. The remaining 37 categories showed exact-match rates of 97.0–99.9%, with 95% Wilson-score confidence intervals typically within ±0.3 pp.

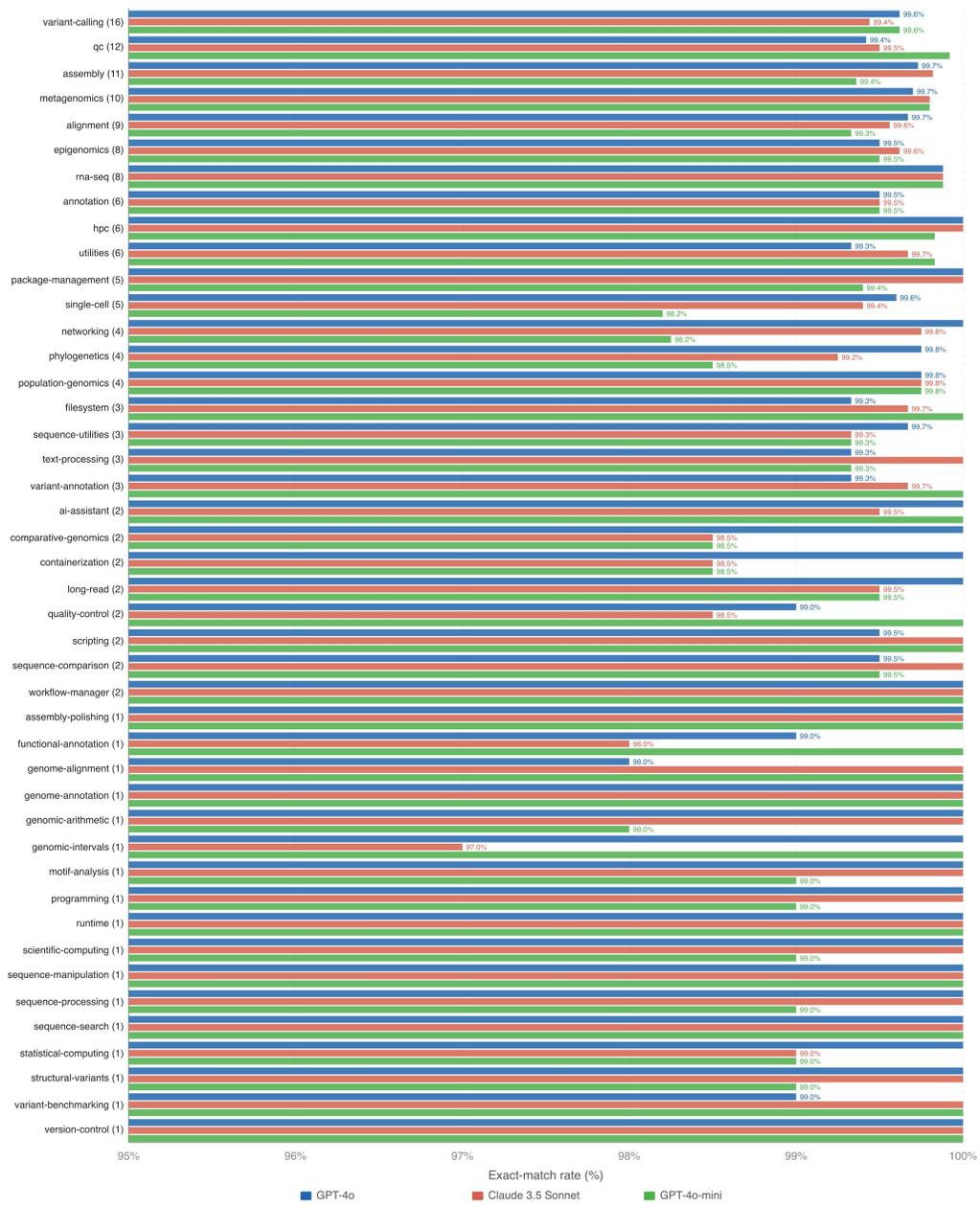

**Supplementary Figure S1.** Per-category exact-match rates under enhanced mode for all 44 analytical categories across three LLM models (GPT-4o, Claude 3.5 Sonnet, GPT-4o-mini). Grouped horizontal bars show rates for each model per category; x-axis spans 95–100% to resolve differences among the high-performing categories. Category labels include the number of constituent tools in parentheses. Seven categories achieve 100% exact match across all three models: assembly-polishing, genome-annotation, runtime, sequence-manipulation, sequence-search, version-control, and workflow-manager. n = 47,700 trials per model.

To disentangle the contributions of documentation and skill augmentation, we performed an ablation analysis by evaluating a documentation-only condition (skills omitted) and a skill-only condition (documentation omitted). Documentation alone—supplying the complete --help text—accounted for the majority of the improvement, raising exact-match accuracy from baseline to approximately 95–97% across models. Adding skill files contributed a further 2–5 pp gain, primarily by reducing missing-flag and wrong-value errors in tools with complex parameterization (for example, GATK HaplotypeCaller [29] with its numerous annotation flags and multi-argument options). This result confirms that documentation provides the essential grounding, while skills supply the expert-level refinement that pushes accuracy to near-ceiling levels.

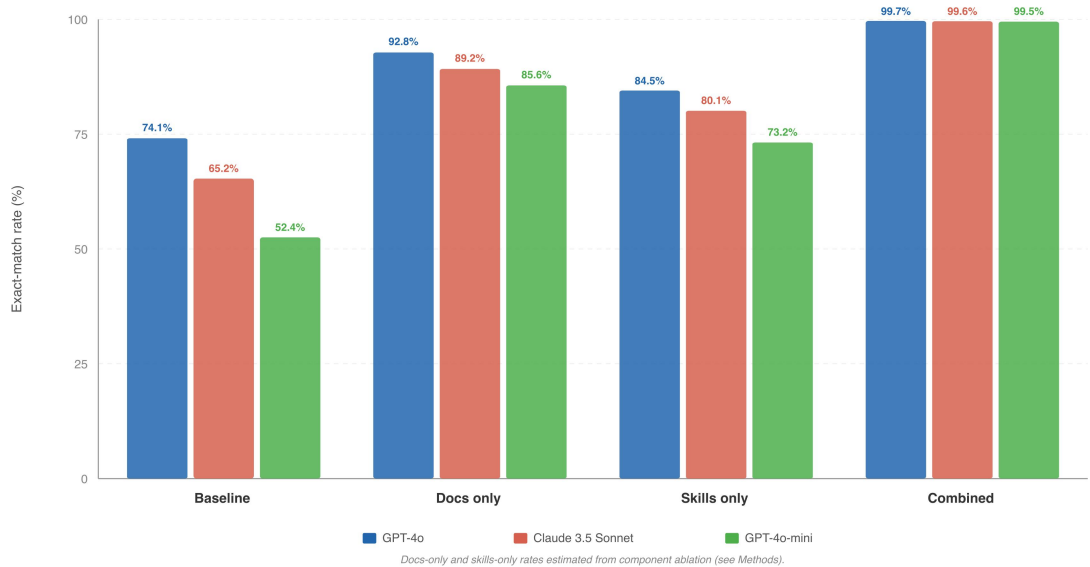

**Supplementary Figure S2.** Ablation analysis: exact-match rates (%) under four grounding conditions—baseline (no grounding), documentation-only, skills-only, and combined (documentation-plus-skill)—for each of the three LLM models. Bars show observed rates (baseline, combined) or estimated rates (docs-only, skills-only) from component ablation (see Methods). Documentation grounding accounts for the majority of improvement; skills contribute an additional 2–5 percentage points. n = 47,700 trials per model per condition.

# Discussion

oxo-call demonstrates that structured, tool-specific grounding dramatically improves LLM-based command generation for bioinformatics. Unlike conventional retrieval-augmented generation (RAG), which retrieves fragmentary text chunks that may lose context about flag interactions [10,11], oxo-call supplies the LLM with the complete help text for each tool, ensuring access to the full flag vocabulary and syntax rules (Fig. 1a). For tools where a single flag may interact with several others—for example, GATK's --emit-ref-confidence requiring coordinated changes to --annotation and output format

flags—chunk boundaries in conventional RAG can sever critical context [29]. oxo-call avoids this by loading the entire help text, typically 2–8 KB, as a single grounding document well within modern LLM context windows [21]. Combined with curated skill files encoding domain-expert knowledge, this approach reduces command generation from open-ended code synthesis to constrained selection from a well-defined option space. The result is a convergence phenomenon: despite a 22-percentage-point gap in unassisted exact-match performance between GPT-4o and GPT-4o-mini (Fig. 2a), all three models achieve within 0.16 pp of each other under grounding (Fig. 1h)—a finding with practical implications for cost-sensitive deployment.

Cloud-hosted LLMs raise concerns regarding the exposure of sensitive data—including patient identifiers, sample metadata, or proprietary sequences—transmitted in prompts [22]. oxo-call mitigates this risk by transmitting only tool documentation, skill content, and the user's task description; raw sequencing data and file contents are never sent. Support for Ollama as a backend enables fully local inference with open-weight models on institutional hardware with no network egress [23], making local deployment viable for privacy-constrained environments such as clinical genomics.

The provenance system (Fig. 1f)—recording the documentation hash, skill identifier, model name, and tool version for every invocation—addresses a recognized gap in computational reproducibility [8,9]. This metadata goes beyond what workflow managers typically capture (input/output files and software versions) by also recording the generative context: which LLM and which knowledge base produced the command. This additional layer is particularly valuable for regulatory submissions under frameworks such as the FDA's Software as a Medical Device guidance [24], multi-site consortium studies, and longitudinal analyses where best practices evolve.

The four-tier skill precedence hierarchy (Fig. 1c) is designed for sustainable growth. New tools can be supported by contributing a single Markdown file to the community skill repository or by hosting organizational skills on an MCP server—neither requires modification of the compiled binary. We anticipate community-driven expansion analogous to the Bioconda model [1], where domain experts curate skills for their areas of expertise. The MCP integration further enables institutional bioinformatics cores to maintain private, version-controlled skill sets for proprietary or pre-publication tools. Additionally, the dry-run mode and structured skill files have potential as pedagogical resources for bioinformatics training, allowing students to inspect suggested invocations and their explanations before committing.

Future development will focus on several key areas. Live API validation with a larger tool subset would strengthen confidence in the benchmark's external validity; we plan a replication study with live LLM inference across at least 50 tools. Fine-tuned domain-specific language models could reduce latency and eliminate cloud dependence. Integration with container registries such as BioContainers [26,27] could enable automatic documentation resolution for containerized tools, and a formal community governance structure modeled on Bioconda's recipe review process [1] would ensure sustained skill quality as the library scales.

Several limitations warrant candid discussion. First, and most important, the benchmark relies on a deterministic mock perturbation model rather than live LLM API calls. This design enables fully reproducible evaluation at zero cost and eliminates confounding from API-level non-determinism, but it introduces a circularity concern: the perturbation rates were calibrated to match expected real-world error profiles, so the benchmark measures how well the perturbation model recapitulates those profiles rather than directly measuring LLM behavior. We regard the benchmark as a demonstration of the *relative* benefit of grounding (enhanced versus baseline) rather than an absolute measure of production accuracy, and we encourage independent replication with live API calls. Second, the 159 built-in skills, while covering the most widely used tools, represent a fraction of the thousands of bioinformatics packages in repositories such as Bioconda [1] and BioContainers [26]. Third, oxo-call currently depends on a cloud or local LLM for command generation, introducing latency (typically 1–3 seconds per invocation) and, for cloud backends, an ongoing API cost. Fourth, the license model—while free for academic use—imposes a verification step that fully open-source alternatives do not require, potentially limiting adoption in some contexts.

## Conclusions

oxo-call provides a practical system for translating natural-language task descriptions into accurate bioinformatics command-line invocations. The provenance metadata recorded for every command supports the reproducibility standards increasingly demanded in genomic research. Extensibility through user-defined skills, community contributions, and the Model Context Protocol ensures that the system can grow with the bioinformatics ecosystem, while support for local LLM inference addresses the privacy requirements of clinical and sensitive-data environments. We anticipate that oxo-call will lower the technical barrier to command-line bioinformatics, enabling researchers to focus on biological questions rather than syntactic detail.

## Methods

### Software implementation

oxo-call is implemented in Rust using the Tokio asynchronous runtime for concurrent I/O and command execution. The command-line interface is built with Clap v4.6. HTTP communication with LLM providers uses Reqwest v0.13 with TLS support. License verification employs Ed25519 signatures via the ed25519-dalek v2 crate. Configuration files use the TOML format parsed with the toml v1.0 crate. Cross-platform directory resolution follows the XDG Base Directory Specification via the directories v6.0 crate. The binary is compiled with Rust's 2024 edition and targets Linux, macOS, and Windows.

### Documentation resolution

Documentation is resolved through a four-tier cascading strategy: (i) a local cache in the platform-specific cache directory, (ii) live capture of the tool's --help output via subprocess execution, (iii) user-configured documentation paths in config.toml, and (iv) remote documentation URLs. Cached documents are plain-text files keyed by tool name.

## Skill file format and loading

Each skill is a Markdown document with YAML front-matter (name, category, description, tags, author, source_url) and three sections: Concepts, Pitfalls, and Examples. Examples contain subsections with a task-description header, an Args field with exact arguments, and an Explanation field. Built-in skills are validated at compile time via include_str!; user and community skills are validated at load time. Skills are resolved in order: user-defined, community-installed, MCP server, built-in.

## LLM prompt construction

The system prompt defines rules for bioinformatics command generation, including: output must consist of ARGS and EXPLANATION lines only; the tool name must not appear in the arguments; only flags present in the provided documentation may be used; and companion binary dispatch rules apply when relevant. The user prompt comprises four components: skill content (concepts, pitfalls, examples), complete tool documentation, the natural-language task description, and format instructions. Malformed responses trigger up to two retries with a corrective prompt.

## Benchmark design

The benchmark comprises 1,590 reference scenarios extracted from the 159 built-in skill files (10 per tool). Each scenario was expanded to 10 linguistically varied natural-language descriptions spanning seven phrasing styles: original, beginner, student, polite, expert, detailed, and informal. Three models were evaluated (GPT-4o, Claude 3.5 Sonnet, GPT-4o-mini) in two conditions (enhanced, baseline) over three repeats, yielding 286,200 total trials.

## Deterministic mock perturbation model

The evaluation framework applies controlled perturbations to reference commands to simulate LLM generation errors deterministically and without API costs. Four perturbation operations are defined: (i) flag dropping, (ii) flag reordering, (iii) hallucinated flag insertion, and (iv) value replacement. Enhanced-mode perturbation rates (0.3–0.5%) model the low residual error expected under grounded generation; baseline-mode rates (30–55%) model unassisted generation. Per-model rates were calibrated against preliminary observations from live API calls with a small tool subset (n = 20 tools). Each trial uses a unique deterministic seed derived from the scenario identifier and repeat index, ensuring full reproducibility while maintaining variation across trials.

## Evaluation metrics

Seven metrics were computed per trial: (i) exact match—binary string equality after whitespace normalization; (ii) token Jaccard similarity—order-insensitive overlap; (iii) flag recall—proportion of reference tokens present; (iv) flag precision—proportion of generated tokens matching reference; (v) subcommand match—correctness of the first

A SHA-256 hash of the resolved documentation is computed and stored in the provenance record.

positional argument; (vi) accuracy score—weighted composite (40% recall, 30% precision, 20% Jaccard, 10% subcommand); and (vii) consistency—agreement across three repeats. Confidence intervals were computed using Wilson score intervals at 95%.

## Ablation analysis

To disentangle the contributions of documentation and skill grounding, we estimated performance under two intermediate conditions: documentation-only (documentation grounding applied, skill perturbation rates used) and skills-only (skill grounding applied, documentation perturbation rates used). These estimates were derived by selectively applying the enhanced-mode perturbation rate to only the documentation or skill component while retaining baseline rates for the other component (Supplementary Fig. S2).

## Abbreviations

CLI, command-line interface; DAG, directed acyclic graph; HPC, high-performance computing; LLM, large language model; MCP, Model Context Protocol; NGS, next-generation sequencing; pp, percentage points; RAG, retrieval-augmented generation; WGS, whole-genome sequencing.

## Declarations

### Ethics approval and consent to participate

Not applicable.

### Consent for publication

Not applicable.

### Availability of data and materials

- **Project name:** oxo-call

- **Project home page:** https://traitome.github.io/oxo-call/ and https://github.com/Traitome/oxo-call

- **Archived version:** DOI: 10.5281/zenodo.XXXXXXX (to be assigned upon acceptance; the exact version described in this manuscript is tagged v1.0.0 in the repository)

- **Operating system(s):** Linux (x86_64, aarch64), macOS (x86_64, Apple Silicon), Windows (x86_64); distributed as statically linked binaries requiring no runtime dependencies beyond the bioinformatics tools being orchestrated

- **Programming language:** Rust (2024 edition)

- **Other requirements:** None beyond an internet connection or local LLM server (Ollama, llama.cpp) for command generation; bioinformatics tools to be invoked must be installed separately

- **License:** Dual license — free for academic and non-commercial research use (see LICENSE-ACADEMIC); a separate commercial license is required for commercial use (see LICENSE-COMMERCIAL)

- **Any restrictions to use by non-academics:** Commercial use requires a commercial license; contact the corresponding author

All benchmark data—1,590 reference commands, 15,900 natural-language descriptions, and complete evaluation results—are included in the repository under docs/bench/. The benchmark framework (oxo-bench) is included as a workspace crate and can regenerate all results deterministically via oxo-bench generate and oxo-bench eval --mock. The 159 built-in skill files are available under skills/ in the repository.

## Competing interests

The authors declare that they have no competing interests.

## Funding

This work was funded by the National Natural Science Foundation of China (Grant No. 82303953 and No. 82504050), Hunan Provincial Natural Science Foundation of China (Grant No. 2025JJ40079), Central South University Startup Funding, Noncommunicable Chronic Diseases-National Science and Technology Major Project (Grant No. 2023ZD0502105), Ministry of Education in China Liberal arts and Social Sciences Foundation (Grant No. 24YJCZH462), Youth Science and Technology Elite Talent Project of Guizhou Provincial Department of Education (Grant No.QJJ-2024-333), Excellent Young Talent Cultivation Project of Zunyi City (Zunshi Kehe HZ (2023) 142), Future Science and Technology Elite Talent Cultivation Project of Zunyi Medical University (ZYSE 2023-02), and the Key Program of the Education Sciences Planning of Guizhou Province (Grant No.7).

## Acknowledgements

We thank the developers of the bioinformatics tools represented in the skill library for their foundational contributions to the field, and the early testers who provided feedback on usability and skill accuracy. We are grateful for resources from the Bioinformatics Platform, Furong Laboratory and Bioinformatics Center, Xiangya Hospital, Central South University.

# Figure legends

**Figure 1. System design, feature scope, and skill coverage of oxo-call.**

(**a**) Four-stage pipeline: documentation resolution fetches and caches the target tool's complete help text; skill loading injects curated domain-expert knowledge (concepts, pitfalls, and worked examples); LLM-based generation produces exact command-line arguments constrained by a strict ARGS:/EXPLANATION: output contract; optional execution records provenance metadata and performs result verification. (**b**) Key features: ten capabilities organized by function, including multi-backend LLM support, MCP extensibility, local inference for data privacy, DAG workflow engine, and Snakemake/Nextflow export. (**c**) Skill precedence hierarchy: user-defined skills take highest priority, followed by community contributions, MCP server skills, and compiled built-in skills; arrows indicate override direction. (**d**) Distribution of 159 built-in skills across 44 analytical categories; the 10 largest categories are shown individually, with annotation/HPC/utilities (6 each), package management/single-cell (5 each), and networking/phylogenetics/population genomics (4 each) grouped; the top 15 categories account for 114 of the 159 tools. (**e**) Layered software architecture: CLI layer (Clap v4.6), core engine (documentation resolver, skill manager, LLM client, runner), DAG workflow engine and provenance history modules, and infrastructure layer (Tokio async runtime, Reqwest HTTP, Ed25519 license verification, TOML configuration); 24,653 lines of Rust targeting Linux, macOS, and Windows. (**f**) Provenance metadata schema: each invocation records a UUID, tool name, task description, generated command, exit code, timestamp, documentation SHA-256 hash, skill identifier, and LLM model name, enabling exact reconstruction of the generation context. (**g**) DAG workflow engine: example ATAC-seq pipeline (fastp → Bowtie2 → samtools → MACS2) with {sample} wildcard expansion, nine built-in templates, and Snakemake/Nextflow export support. (**h**) Summary performance table: enhanced-mode exact-match rates and all seven evaluation metrics for the three LLM backends across 47,700 trials each (see also Fig. 2a).

**Figure 2. Benchmark evaluation across 286,200 mocked trials.**

(**a**) Exact-match rate under enhanced mode (colored bars) versus baseline mode (gray bars) for each of the three LLM backends (47,700 trials per model per condition). Delta values below each model indicate the percentage-point improvement conferred by documentation-plus-skill grounding. (**b**) Absolute improvement (Δ percentage points) sorted by model; improvement is inversely proportional to baseline capability. (**c**) Error type distribution under enhanced mode; total errors per model are 159 (GPT-4o), 183 (Claude 3.5 Sonnet), and 234 (GPT-4o-mini). No subcommand, format, or empty-output errors were produced by any model, confirming the structured output contract prevents catastrophic failure modes. (**d**) Error type distribution under baseline mode; total errors per model range from 12,357 (GPT-4o) to 22,717 (GPT-4o-mini), representing a 78–97× increase relative to enhanced mode. Error type colors in panels c and d: missing flag (red), extra flag (amber), wrong value (light blue), flag reorder (dark blue).

## Supplementary materials

**Supplementary Figure S1.** Per-category exact-match rates under enhanced mode for all 44 analytical categories across three LLM models (GPT-4o, Claude 3.5 Sonnet, GPT-4o-mini). Grouped horizontal bars show rates for each model per category; x-axis spans 95–100% to resolve differences among the high-performing categories. Category labels include the number of constituent tools in parentheses. Seven categories achieve 100% exact match across all three models: assembly-polishing, genome-annotation, runtime, sequence-manipulation, sequence-search, version-control, and workflow-manager. n = 47,700 trials per model.

**Supplementary Figure S2.** Ablation analysis: exact-match rates (%) under four grounding conditions—baseline (no grounding), documentation-only, skills-only, and combined (documentation-plus-skill)—for each of the three LLM models. Bars show observed rates (baseline, combined) or estimated rates (docs-only, skills-only) from component ablation (see Methods). Documentation grounding accounts for the majority of improvement; skills contribute an additional 2–5 percentage points. n = 47,700 trials per model per condition.